\documentclass[aps,prb,reprint,floatfix,superscriptaddress]{revtex4-2}

\usepackage{graphicx}
\usepackage{amsmath,amssymb,amsfonts}
\usepackage{url,hyperref}
\usepackage{multirow}
\usepackage[version=4]{mhchem}
\usepackage{subcaption}
\usepackage{verbatim}
\usepackage{titlesec}
\usepackage{enumitem}
\usepackage{color}
\usepackage[utf8]{inputenc}

\usepackage{xcolor}
\usepackage{soul}
\sethlcolor{white}

\graphicspath{{/}}

\begin{document}
\title{Structural Dynamics Descriptors for Metal Halide Perovskites}

\author{Xia Liang}
\affiliation{Department of Materials, Imperial College London, South Kensington Campus, London SW7 2AZ, UK\\}

\author{Johan Klarbring}
\affiliation{Department of Materials, Imperial College London, South Kensington Campus, London SW7 2AZ, UK\\}
\affiliation{Department of Physics, Chemistry and Biology (IFM), Link\"{o}ping University, SE-581 83, Link\"{o}ping, Sweden}

\author{William Baldwin}
\affiliation{Department of Engineering, University of Cambridge, Cambridge CB2 1PZ, UK}

\author{Zhenzhu Li}
\affiliation{Department of Materials, Imperial College London, South Kensington Campus, London SW7 2AZ, UK\\}

\author{G\'abor Cs\'anyi}
\affiliation{Department of Engineering, University of Cambridge, Cambridge CB2 1PZ, UK}

\author{Aron Walsh}
\email{a.walsh@imperial.ac.uk}
\affiliation{Department of Materials, Imperial College London, South Kensington Campus, London SW7 2AZ, UK\\}
\affiliation{Department of Physics, Ewha Womans University, Seoul 03760, Korea}

\date{\today}

\begin{abstract}
\begin{center}
    \textbf{Abstract}
\end{center}
Metal halide perovskites have shown extraordinary performance in solar energy conversion technologies. They have been classified as ``soft semiconductors" due to their flexible corner-sharing octahedral networks and polymorphous nature. Understanding the local and average structures continues to be challenging for both modelling and experiments. Here, we report the quantitative analysis of structural dynamics in time and space from molecular dynamics simulations of perovskite crystals. \hl{The compact descriptors provided cover a wide variety of structural properties}, including octahedral tilting and distortion, local lattice parameters, molecular orientations, as well as their spatial correlation. To validate our methods, we have trained a machine learning force field (MLFF) for methylammonium lead bromide (\ce{CH3NH3PbBr3}) using an on-the-fly training approach with Gaussian process regression. The known stable phases are reproduced and we find an additional symmetry-breaking effect in the cubic and tetragonal phases close to the phase transition temperature. To test the implementation for large trajectories, we also apply it to 69,120 atom simulations for \ce{CsPbI3} based on an MLFF developed using the atomic cluster expansion formalism. The structural dynamics descriptors and Python toolkit are general to perovskites and readily transferable to more complex compositions. 

\begin{center}
    \textbf{TOC Graphic} 
\end{center}
\begin{center}
    \includegraphics[width=7cm]{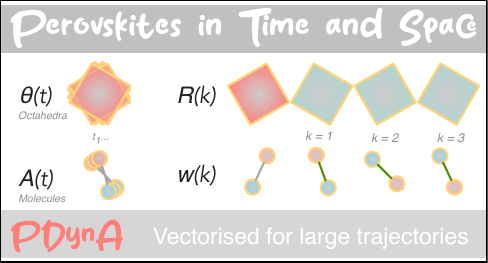}
\end{center}
\end{abstract}

\maketitle

\section{Introduction}

Perovskite is one of the most important classes of crystal structure in materials chemistry owing to the large number of accessible compositions and variety of structural derivatives. 
For standard ABX$_{3}$ compounds, the A-site cation is positioned inside a cuboctahedral cage formed of corner-sharing BX$_{6}$ octahedra. When the X-site is occupied by a halide, the A cation is monovalent and the B cation is divalent, the material is called a metal halide perovskite.
\hl{The term hybrid organic-inorganic halide perovskite can refer to compounds where the A-site is occupied by a molecular cation,} e.g. formamidinium (FA$^+$) or methylammonium (MA$^+$), while the B-site cation is an inorganic cation such as Pb$^{2+}$ or Sn$^{2+}$.
These perovskites have become a popular research topic in the context of photovoltaics and related optoelectronic technologies.\cite{perov_sc_long_review1}

The cubic perovskite archetype structure (space group Pm3m), which features a corner-sharing array of regular octahedra, is typically only observed at higher temperatures. 
A series of lower symmetry perovskite phases emerge at lower temperatures. 
These polymorphs can be classified according to the average octahedral tilting patterns\cite{glazer_original,howard1998group,woodward1997octahedral} and to the types of ion displacements that occur\cite{stokes2002group}. 
The instantaneous structure often differs from the structure averaged over space and time, which can be seen from differences between X-ray diffraction measurements and more local probes such as electron diffraction,\cite{baikie2013synthesis} X-ray scattering,\cite{beecher2016direct} Raman spectroscopy,\cite{yaffe2017local} and nuclear magnetic resonance\cite{kubicki2020local}. 

There are advantages to using hybrid perovskites in photovoltaic devices, such as long photogenerated carrier diffusion lengths, which have been related to the polar nature of the organic A-site cations.\cite{stranks2013electron,aron_origin_performance} However, the presence of the molecular species introduces an additional degree of rotational freedom in the crystal. The structure of the light elements (e.g. C, H, N) can be difficult to distinguish from the heavy frameworks (e.g. Pb, Br, I) using standard characterisation techniques such as X-ray diffraction. The molecular arrangement and reorientations have been probed from techniques including quasi-elastic neutron scattering\cite{leguy2015dynamics} and time-resolved infrared spectroscopy\cite{bakulin2015real}. Compositional engineering of perovskites to produce stable and high-performant materials has led to the study of such complex mixtures as 
Cs$_{0.05}$FA$_{0.78}$MA$_{0.17}$Pb(I$_{0.83}$Br$_{0.17}$)$_{3}$.\cite{triple_cation1} Experimental observations on these state-of-the-art materials suggest that the (average) cubic perovskite phase of this material possesses (local) symmetry-breaking tilting of octahedra, that hinders degradation pathways.\cite{perov_tilting_stable_science} 

From the materials simulation perspective, modelling of perovskite crystal dynamics at finite temperatures can be achieved using molecular dynamics (MD).\cite{quarti2015structural,mattoni2016modeling,egger2016hybrid,dar2016origin,stoich_MD_perov_mixed_anion_cation,guo2022atomistic} 
However, large and long simulations are required to quantify local fluctuations in atomic positions and correlations, for example, between rotations of the molecular sublattice and titling of the octahedral inorganic networks.  
MD simulations based on density functional theory (DFT) forces are restricted in system size (typically 10s---100s of atoms for 100s ps) due to the high computational cost, and conventional force fields are lacking in their ability to describe both the dynamics of the organic A-site cations and the highly deformable octahedra. Alternative methods are necessary to enable the collection of accurate and efficient structural information. 

\subsection{Machine learning force fields}

Traditional empirical descriptions of interatomic interactions rely on a fixed functional form, such Lennard-Jones or Buckingham potentials.\cite{lewis1985potential}
Machine learning force fields (MLFF) involve a more flexible model to learn and predict quantities such as energy, atomic forces and stresses (EFS) of crystal structures from an input (training) dataset.\cite{ceriotti2021introduction} 
MLFFs are promising to reach a DFT-level accuracy with an efficiency comparable to classical potentials. In such approaches, the atomic structures are converted into a symmetrized feature space and functions on these spaces thus can be learned with general ML regression algorithms. 

Gaussian process regression is a powerful tool for non-linear function approximation in high-dimensional parameter spaces.\cite{RasmussenGP}
It is used in the Gaussian approximation potential (GAP) formalism, where the local environments of atoms in a crystal structure are converted into a \hl{structural feature} space.\cite{gap_original,deringer2021gaussian} 
It is commonly used in combination with a local atom-centered \hl{representation} called Smooth Overlap of Atomic Positions (SOAP)\cite{soap}, which incorporates two-body radial and three-body angular contributions. 
The reference configurations for force field training can be selected in several ways, including through the use of Bayesian error estimation. Based on this approach, Jinnouchi \textit{et al.}~\cite{vaspmlff_original} implemented an on-the-fly machine learning force field method with DFT-level accuracy in predicting energy and forces of atomic configurations and is \textit{ca.} 100 times faster than standard DFT calculations. There have been prior successes in the application of this approach to metal halide perovskites.~\cite{vaspmlff_other_mapb_halides}

\subsection{Perovskite structural descriptors}
The temperature dependence of the unit cell parameters is a commonly tracked feature of crystallographic perovskite phases. However, a quantitative description of the underlying octahedral deformations and the A-site cation dynamics can reveal more important details. 

The common language for classifying octahedral tilting is the Glazer notation~\cite{glazer_original}, where three-dimensional tilting can be described with three magnitudes, each denoting the tilting about $a$, $b$ and $c$ axes, and the corresponding in-phase(+)/anti-phase(-) tilting correlation pattern along that direction. For instance, the Glazer notation of an orthorhombic perovskite phase can be $a^{+}b^{-}b^{-}$, meaning that the tilt angles about $b$ and $c$ axes are the same and different from that of $a$ axis, and the tilting about $a$ axis is in-phase and the other two are anti-phase. 
Octahedral tilting can also be described using the B--X--B bond angle~\cite{bxb_angle} or X--B--B--X dihedral angle~\cite{xbbx_dihedral}. 

Beyond the analysis of static or globally averaged structures (coordinates averaged in time and space), the temporal and spatial behaviour of perovskite structures is of growing importance for understanding inhomogeneities in perovskite materials and performance bottlenecks in perovskite devices.\cite{tennyson2019heterogeneity} 

Here, we developed the Perovskite Dynamics Analysis (\textsc{PDynA}) software package to systematically quantify the dynamic behaviour of perovskite materials in an automatic and efficient manner. The integrated analysis tools aim to create an intuitive picture of the structural dynamics, as well as to provide \hl{compact descriptors} for future data-driven studies of perovskites. Such analysis can provide validation for machine learning force fields and systematically interpret MD trajectories. The package combines standard quantification methods with novel descriptors to build a full view of the structural dynamics in perovskite crystals. 

\begin{figure}[h]
    \centering
    \includegraphics[width=1.0\linewidth]{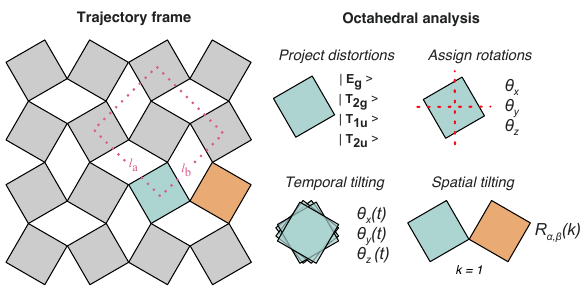}
\caption{
\hl{Illustration of the procedure for structural information extraction from a molecular dynamics trajectory frame of an ABX$_3$ perovskite including the definition of local lattice spacing from the relative positions of the B atoms, and the assignment of BX$_6$ octahedral tilting.}
}
    \label{structures}
\end{figure}

\section{Methods} 

\subsection{Perovskite dynamics analysis}
\textsc{PDynA} is an integrated Python code that is tested with Python 3.8+. 
The algorithm first identifies all constituent octahedra and their connectivity as well as the A-site cations in the perovskite structure. \hl{We then formulate a set of compact descriptors to represent the structural characteristics in space and time, which can be compared for different external conditions and chemical compositions.}

Functions are developed to analyse MD trajectories in four general aspects: lattice parameters (relative position of BX$_6$ octahedra), octahedral tilting, octahedral distortion, and A-site molecular orientation and displacement. 
The computation of these properties is vectorised with multi-threading to enable faster processing of large trajectories. 
The analysis of dynamic structural behaviour can be done in two modes, static (time-independent) analysis mode and transient analysis mode (time-dependent or temperature-dependent). 
In the static mode, the behaviour of constant-temperature MD trajectory will be sampled after a given equilibration period and output static properties in the form of cumulative data. The sampling frequency is automatically determined according to the system size. In the transient mode, computed properties will be sampled evenly across the trajectory and provide structural information with respect to time or temperature. 

In this work, $a$, $b$ and $c$ are the principal axes related to crystallographic perovskite phases. These three axes have a one-to-one correspondence to the principal Cartesian axes of the sample, $x$, $y$ and $z$ unless specified otherwise.  Under most circumstances, these two systems have one-to-one correspondence except during a phase transition event (e.g. when cooling from a cubic to a tetragonal phase, a tetragonal distortion may emerge along $x$, $y$ or $z$ of the simulation cell).
We use $\alpha$ and $\beta$ to denote variable symbols associated with one of the sample-specific axes.  

\textbf{Lattice parameters}: Lattice parameters of the unit cell are straightforward indicators of crystallographic phases for perovskite materials. These can be defined 
globally from the dimensions of the supercell or locally from the atomic positions in the MD trajectory. 
We calculate three local lattice spacings $l_{a}$, $l_{b}$ and $l_{c}$ from the relative pairwise positions of the B-site cations following Ref. \onlinecite{vaspmlff_perov} and illustrated in 
Fig.~\ref{structures}.
These local spacings can be converted into three (pseudo-cubic) local lattice parameters $l_{pc,a}$, $l_{pc,b}$ and $l_{pc,c}$ with the relationships $l_{pc,a} = \frac{l_{a}}{\sqrt{2}},\: l_{pc,b} = \frac{l_{b}}{\sqrt{2}},\: l_{pc,c} = \frac{l_{c}}{2}$. 

\textbf{Octahedral distortions}: \hl{An obstacle to the analysis of perovskite structures is that the BX$_6$ octahedral building blocks are not rigid but deform in time.
Group theory can be applied to construct a quantitative and compact measure of octahedral distortions}.
Magnitudes of the four irreducible distortion modes, namely $E_{g}$, $T_{2g}$, $T_{1u}$, $T_{2u}$, are computed for each octahedron in space and time, following the procedure outline by Morita et al.\cite{kazuki_distortion}  
The function uses
the Hungarian order matcher method~\cite{hungarian_method} as implemented in \textsc{pymatgen}~\cite{pymatgen_original}.
Gaussian peaks are fitted to each population to determine the mean and standard deviation of the dynamic distortion behaviour of the structure. Also, the octahedral distortions of the \textit{time-averaged} structure can be calculated to analyse static global distortions in the phase of interest.  

\textbf{Octahedral tilting}: \hl{Together with the identification and removal of distortion modes, each octahedron is matched to an ideal reference configuration.} We created a function to simultaneously analyse the octahedral tilting and distortions and output the three-dimensional tilting status as a rotation matrix.
This method follows the resolved connectivity of B and X-sites, and avoids anomalies arising from the structural mismatch, differing from the implementation used in Ref. \cite{kazuki_distortion}. 

We denote the \hl{individual} tilting angle of an octahedron around axis $\alpha$ at time $t$ by:
\begin{equation}
\label{eq:oct_tilt}
\theta_\alpha(t;\mathbf{n}),
\end{equation}
where $\mathbf{n} = (n_x,n_y,n_z)$ are 3 integer coordinates indexing the octahedra in the supercell. We identify the tilting angles in Eq.\ \ref{eq:oct_tilt} with the three Euler rotation angles computed from the rotation matrix. By symmetry, $\theta_\alpha(t;\mathbf{n})$ takes the range $-45^{\circ} < \theta_{\alpha} < 45^{\circ}$. 
The tilting of octahedra is spatially correlated between neighbours, a crucial feature in the understanding of perovskite phases. 
\hl{We track the tilting angle relationship between an octahedron \textbf{n} and its $k$th neighbour in the $x$ direction using the expression}:
\begin{equation}
  r_{\alpha,x}^{(k)}(t;\mathbf{n}) =  \frac{\theta_\alpha(t;n_x,n_y,n_z)\theta_\alpha(t;n_x+k,n_y,n_z)}{\sqrt{|\theta_\alpha(t;n_x,n_y,n_z)\theta_\alpha(t;n_x+k,n_y,n_z)|}}
 \label{eq:r} 
\end{equation}
and similarly for tilts in the $y$ and $z$-directions. We record the distribution of $r_{\alpha,x}^{(k)}(t;\mathbf{n})$ over all octahedra, all timesteps, and over $\pm k$ along an MD trajectory. This distribution can then be plotted as a function of tilt angle (see Figure~\ref{res:tilt}). The sign of $r_{\alpha,\beta}^{(k)}$, $\beta = x, y$ or $z$, reflects the nature of in-phase (positive) or anti-phase (negative) correlation. For the special case of $\alpha=\beta$ and $k=1$, the correlation reflects the first nearest neighbour tilt angle  along the tilt direction, which is related to the superscript of the Glazer notation. For $\alpha \neq \beta$, $r_{\alpha,\beta}^{(k)}$ reflects the off-axis correlation, i.e., the relationship between tilts of neighbouring octahedra in a plane orthogonal to the tilting direction.  

We further calculate \hl{global} spatial tilting correlation functions as:
\begin{equation}
  R_{\alpha,x}(k) = C\left\langle |\theta_\alpha(t;n_x,n_y,n_z)\theta_\alpha(t;n_x+k,n_y,n_z)| \right\rangle_{\mathbf{n},t, \pm k},
\end{equation}
where, $\langle ... \rangle_{\mathbf{n},t, \pm k}$ denotes an average over all octahedra in the supercell, over simulation time and over neighbours in both the positive and negative $x$ direction and $C$ is a constant such that $R_{\alpha,\beta}(0) = 1 $.  
The spatial extent of the correlation of octahedral tilts can be quantified by fitting $R_{\alpha,\beta}(k)$ to a decaying exponential:
\begin{equation}
    R_{\alpha,\beta}(k)= \exp\left(-\frac{k}{\xi_{\alpha,\beta}}\right),
\end{equation}
where $\xi_{\alpha,\beta}$ is the fitted correlation length in the $\beta$ direction of the tilt around the $\alpha$-axis. $\xi_{\alpha,\beta}$ forms a second rank tensor with nine components. 

We define the tilting correlation polarity (TCP) around direction $\alpha$, $\delta_{\alpha}$, by comparing the positive and negative correlation population counts from $r_{\alpha,\alpha}^{(1)}$ (Eq.~\ref{eq:r})
\begin{equation}
\label{eq:tcp}
    \delta_{\alpha} = \frac{n_{\alpha}^{+}-n_{\alpha}^{-}}{n_{\alpha}^{+}+n_{\alpha}^{-}}
\end{equation}
where $n_{+}$ and $n_{-}$ are the counts of positively ($r_{\alpha,\alpha}^{(1)}$ $>$ 0) and negatively ($r_{\alpha,\alpha}^{(1)}$ $<$ 0) correlated octahedron pairs, respectively. As a result, $\delta_{\alpha}$ ranges from -1 to 1. Values of $\delta_{\alpha}$ close to 1, 0 and -1 correspond to Glazer notation superscripts of $+$, $0$ and $-$, respectively.

\textbf{Molecular orientation}: For hybrid compositions, the molecular components can rotate in three-dimensional space. One or multiple vectors are used to describe the molecular orientation (MO). The molecules currently implemented are MA and FA, but an interface is provided for user-defined molecules. 
For MA (\ce{CH3NH3}) a single vector $\mathbf{v}_{\mathrm{MA}}$, connects the C and N atoms and is sufficient to fully describe the rotation (if the contribution from hydrogen atoms is ignored). 
For FA (\ce{CH(NH2)2}), two vectors are required to fully describe the MO. 
The polarization vector $\mathbf{v}_{\mathrm{FA1}}$ is the vertical bisector from the C atom to the connection between the two N atoms.
The second vector $\mathbf{v}_{\mathrm{FA2}}$ connects the two N atoms. 

We denote the unit length MO of a molecule at time $t$, $\mathbf{v}(t;\mathbf{n})$, where $ \mathbf{n} = (n_x,n_y,n_z)$, again, are 3 integer coordinates indexing the molecules in the supercell.
Akin to the case of octahedral tilting, we track the spatial correlation of MOs between $k$th neighbours in the $x$ direction at time $t$:
\begin{equation}
        w_{x}^{(k)}(t;\mathbf{n}) = \mathbf{v}(t;n_x,n_y,n_z) \cdot \mathbf{v}(t;n_x+k,n_y,n_z),
\end{equation}
and similarly for neighbours in the $y$ and $z$ directions. The distribution of $w_{\alpha}^{(k)}$, $\alpha = x, y$ or $z$, over all octahedra and simulation time is denoted $C_{\alpha}^{(k)}(w)$. 
Based on this distribution, we define two quantitative descriptors for the spatial correlation MOs; the alignment factor (AF) and the contrast factor (CF). 
The AF in the $\alpha$ direction is:
\begin{equation}
\mathrm{AF}_{\alpha} = \frac{\int_{0}^{1} |w| C_{\alpha}^{(k)}(w) dw-\int_{-1}^{0} |w| C_{\alpha}^{(k)}(w) dw}{\int_{-1}^{1} |w| C_{\alpha}^{(k)}(w) dw},
\end{equation}
and quantifies the relative difference between the tendency of aligned ($\int_{0}^{1} |w| C_{\alpha}^{k}(w) dw$) and anti-aligned ($\int_{-1}^{0} |w| C_{\alpha}^{k}(w) dw$) nearest neighbour molecular order. 
The CF measures the similarity of the relative alignment of the first- and second-nearest neighbour in a direction $\alpha$:
\begin{equation}
\mathrm{CF}_{\alpha} = \frac{\int_{-1}^{1} \min\{C_{\alpha}^{(1)}(w),C_{\alpha}^{(2)}(w)\} dw}{\int_{-1}^{1} C_{\alpha}^{(1)}(w) dw}.
\end{equation}
This quantity indicates the similarity between the first and second nearest neighbour distributions, $C_{\alpha}^{(1)}$ and $C_{\alpha}^{(2)}$, by taking the overlapping portion of them as a fraction of the entire population. We then average $\mathrm{CF_{\alpha}}$ over the three Cartesian directions,
\begin{equation}
\mathrm{CF} = \frac{1}{3} \sum_{\alpha} \mathrm{CF_{\alpha}}.
\end{equation}
$\mathrm{CF}$ and $\mathrm{AF}$ take values between in the ranges $[0 ,1]$, and $[-1, 1]$, respectively. 
We choose to rescale $\mathrm{CF} \leftarrow 2(\mathrm{CF}-\frac{1}{2})$ to map it onto $[-1, 1]$ to make plotting and comparison between the two descriptors more convenient. 

We further compute the temporal autocorrelation of the molecular orientation:
\begin{equation}
A_{MO}(t) = \langle \mathbf{v}(t_{0};\mathbf{n}) \cdot \mathbf{v}(t_{0}+t;\mathbf{n}) \rangle_{t_0,\mathbf{n}}
\end{equation}
The characteristic times associated with the reorientation of the MOs are calculated by fitting this autocorrelation to a sum of two exponential decays~\cite{vaspmlff_perov}: 
\begin{equation}
    A_{MO}(t) = C \: \exp\left(-\frac{t}{\tau_{1}}\right) + (1-C) \: \exp\left(-\frac{t}{\tau_{2}}\right)
\label{equ:autocorr}
\end{equation}
where $C$ is a fitted constant for tuning the linear combination of the two exponential functions and $\tau_{1}$ and $\tau_{2}$ are time constants. Typically, $C$ is approximately 0.9 after parameter optimisation, indicating that $\tau_{1}$ is the dominant time scale, while the second exponential term acts as a correction term for the fast initial decay, with typical values of $\tau_{2}$ around 0.5$\,$ps. Therefore, the calculated time constant for the observed decaying autocorrelation function is the fitted parameter $\tau_{1}$. 

In addition to these core functions, \textsc{PDynA} also contains functions that calculate \textit{time-averaged} structure, radial distribution function, A-site cation displacement, composition heterogeneity analysis, three-dimensional visualisation of MO and tilt domain, etc.
The visualisation of atomic structures in this work is performed with VESTA~\cite{vesta}, and Matplotlib~\cite{matplotlib} is used for plotting the rest of the results. 

\begin{figure}[hbt]
    \centering
    \includegraphics[width=1.0\linewidth]{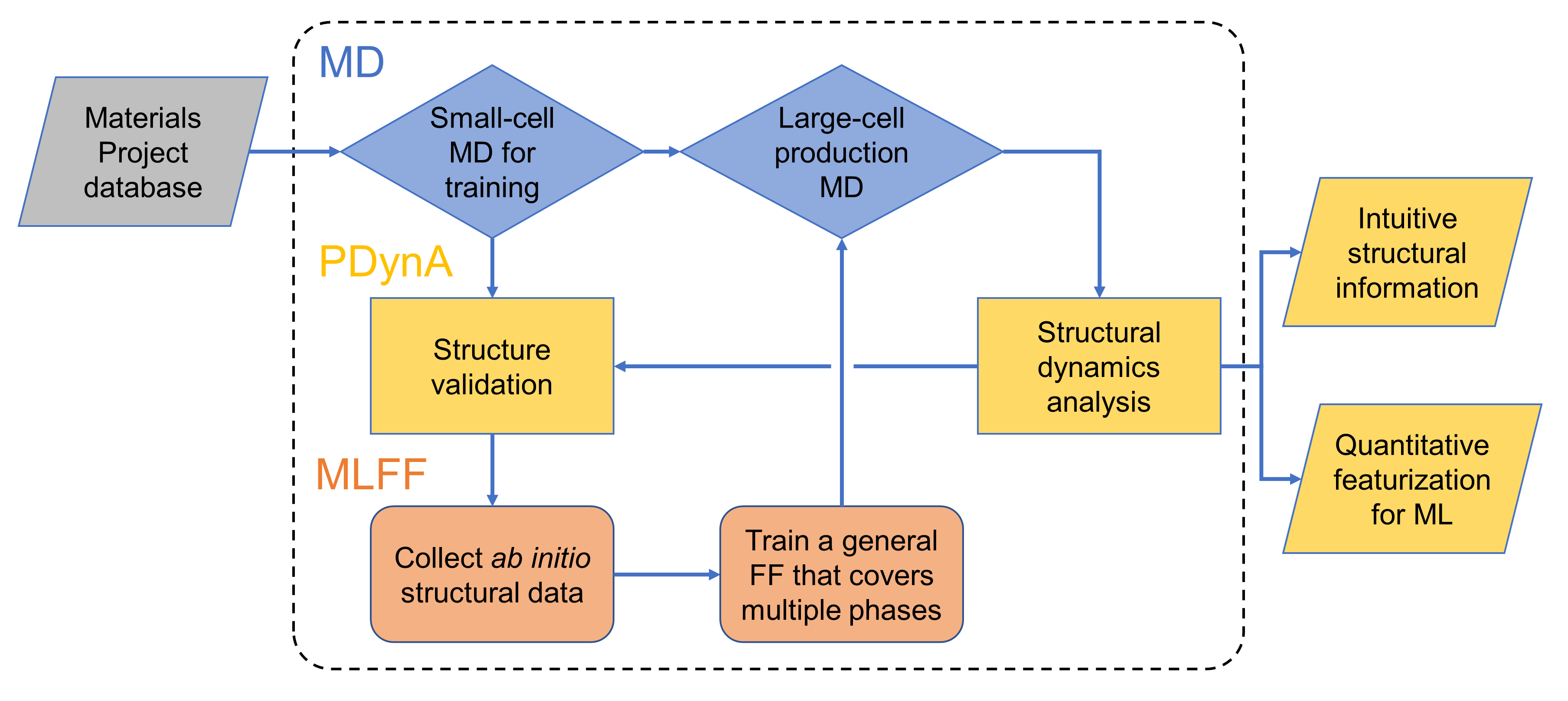}
    \caption{Integrated workflow used for investigating a perovskite compound of interest, combining structural analysis using \textsc{PDynA} of molecular dynamics (MD) trajectories based on a machine learning force field (MLFF). 
    }
    \label{workflow}
\end{figure}

\subsection{MLFF training and trajectories for \ce{CH3NH3PbBr3}}
\label{sec:mlff_for_MAPbBr3}
Molecular dynamics trajectories are generated for \ce{CH3NH3PbBr3} using a custom machine learning force field. 
The MLFF is based on Gaussian process regression with Bayesian error estimation~\cite{vaspmlff_original} as implemented within the Vienna \textit{Ab initio} Simulation Package (\textsc{VASP}).~\cite{vasp_original1,vasp_original2} The energies, forces and stresses used for the training data were calculated based on DFT.
Projector augmented wave (PAW) potentials are used for all DFT calculations. The chosen PAW potentials for the involved elements are Cs\_sv ($5s^{2}5p^{6}6s^{1}$), C ($2s^{2}2p^{2}$), N ($2s^{2}2p^{3}$), H ($1s^{1}$), Pb\_d ($5d^{10}6s^{2}6p^{2}$), I ($5s^{2}5p^{5}$), and Br ($4s^{2}4p^{5}$), all projection operator evaluations are in reciprocal space. 
The r$^{2}$SCAN exchange-correlation functional~\cite{r2scan_original1} is selected. The energy threshold for electronic convergence is set to 10$^{-5}\,$eV, and the plane wave basis set cut-off energy is 500$\,$eV. Gaussian smearing with a width of 50$\,$meV is adopted for the smearing of electronic band occupancy. A $2\times2\times2$ $\Gamma$-centred \textit{k}-point grid is used for all training calculations.

For the MD simulations, an isothermal-isobaric (NpT) ensemble with an external pressure of 1 bar is first adopted to acquire equilibrium lattice parameters and atomic positions on a $2\times2\times2$ supercell (96 atoms) at each temperature of interest, the initial structure is collected from the Material Project database~\cite{matproj1}. This MD calculation is accelerated with the on-the-fly learning MLFF, the local configurations collected in this step are neglected. The equilibrium configuration\hl{(average lattice parameters and atomic positions at that temperature)} is then imported into another isothermal-isochoric (NVT) ensemble with the same supercell size. \textit{Ab initio} data for force field training is retrieved from the NVT MD also using the on-the-fly learning mode. A Langevin thermostat is applied to both ensembles, and atomic and lattice friction constants are set to 10$\,$ps$^{-1}$. The MD time step for hybrid halide perovskites is set to 0.5$\,$fs, and 100,000 steps are performed in each MD calculation. 

Independent on-the-fly MLFF training processes are performed for the constituent perovskite polymorphs at 100 K ($\gamma$), 160 K ($\beta$), 210 K ($\beta$), 350 K ($\alpha$). The on-the-fly learning is achieved through Bayesian error estimation, where a DFT calculation will be performed only if the predicted error is above a threshold.~\cite{vaspmlff_original}. 
The training data picked up from all four on-the-fly MD runs are combined into a final training set used to produce a general force field. The resulting training set contains 1520 DFT snapshots, from which 690, 602, 3729, 780 and 129 local reference configurations for Br, C, H, N and Pb were extracted. 

All MLFF-related parameters are consistent with the defaults of \textsc{VASP} 6.3.0. \hl{The cutoff radius for the radial and angular representations is 5 \AA, and the width of the Gaussian broadening for both representations is 0.5 \AA.} Validation of the general force field is achieved through another set of $6\times6\times6$ supercell (2592 atoms) production runs at each training circumstance, and the structure is analysed similarly to examine if the lattice parameters and global tilting patterns are consistent with the corresponding experimental measurements. Production runs are performed in a $6\times6\times6$ supercell with the same MD parameters as the training session except that the time step is increased from 0.5$\,$fs to 1$\,$fs. The computational workflow is depicted in Fig.~\ref{workflow}.

\subsection{MLFF training and trajectories for \ce{CsPbI3}}

We have demonstrated the scalability of \textsc{PDynA} by analysing large-scale simulations of the inorganic perovskite \ce{CsPbI3}. 
An MLFF based on the Atomic Cluster Expansion (ACE)~\cite{ace2019drautz} framework was used to simulate the material. The ACE formalism generalises the SOAP representation used by the Gaussian approximation potentials discussed in section \ref{sec:mlff_for_MAPbBr3}. 
Specifically, ACE constructs a systematic body-ordered expansion of the potential energy landscape using a linear basis for symmetric polynomials and can be taken to arbitrary body order. The ACE basis is complete in principle, in contrast to SOAP or other three- or four-body representations\cite{incompleteness}. A detailed discussion of ACE can be found in Ref.~\cite{ace2022dusson}.

The ACE model for \ce{CsPbI3} has been presented previously~\cite{baldwin2023dynamic} and is publicly available along with the training dataset. 
MD simulations of $24\times24\times24$ supercells containing 69,120 atoms were performed using LAMMPS\cite{lammps} in the NpT ensemble. The simulation timestep was 4 fs, the system was equilibrated for 1 ns and data was then collected for 1 ns. 

\section{Results and Discussion}

\begin{figure*}[htb]
    \centering
    \includegraphics[width=1.0\textwidth]{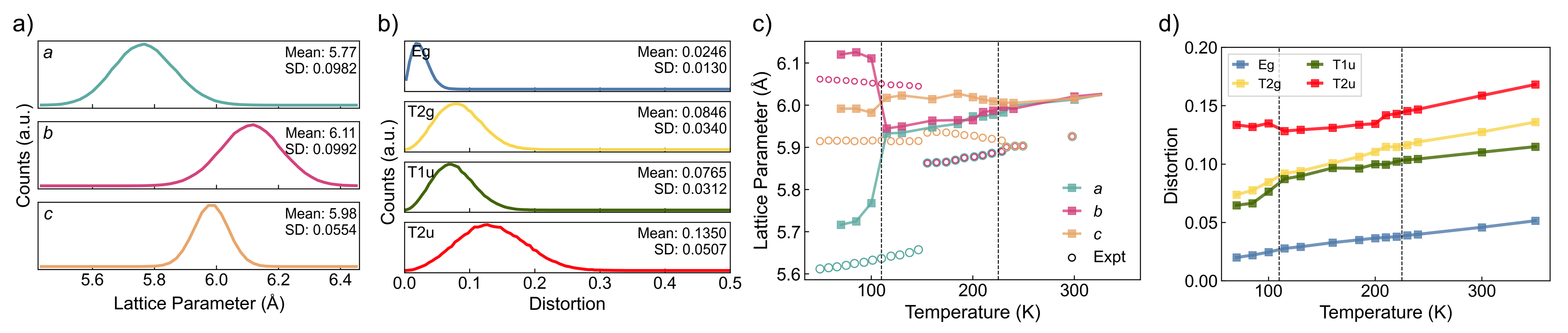}
    \caption{Dynamic distributions of (a) local lattice parameters and (b) octahedral distortions of \ce{MAPbBr3} at 100$\,$K. Gaussian peaks are fitted to each distribution for a mean and standard deviation (SD). (c) Lattice parameters versus temperature. Experimental values\cite{mapbbr3_lattice_ref} are illustrated with circles. (d) The distortion magnitudes versus temperature. In c and d, the phase transition temperatures are indicated with vertical dashed lines.}
    \label{res:gaussians}
\end{figure*}

\subsection{\ce{MAPbBr3}}
\subsubsection{Properties of the ground state} 

The distribution of lattice parameters and octahedral distortions of orthorhombic \ce{MAPbBr3} at 100$\,$K are shown in Fig.~\ref{res:gaussians}(a,b). 
Both properties are evenly distributed around a mean value and can be approximated with a Gaussian distribution. Therefore, the global lattice parameters and distortion modes at constant temperature can be condensed into three and four characteristic values, respectively. The standard deviation of the $c$ lattice parameter is smaller than the other two because the pseudo-cubic convention distinguishes the $c$ value from $a$ and $b$, by giving them different divisors. 
The calculated lattice parameters (Fig.~\ref{res:gaussians}c) are approximately 0.1 \AA ~larger than measured values.\cite{mapbbr3_lattice_ref}

\subsubsection{Temperature Induced Phase-Transformations and Octahedral Tilting} 
\begin{figure}[htb]
    \centering    \includegraphics[width=0.74\linewidth]{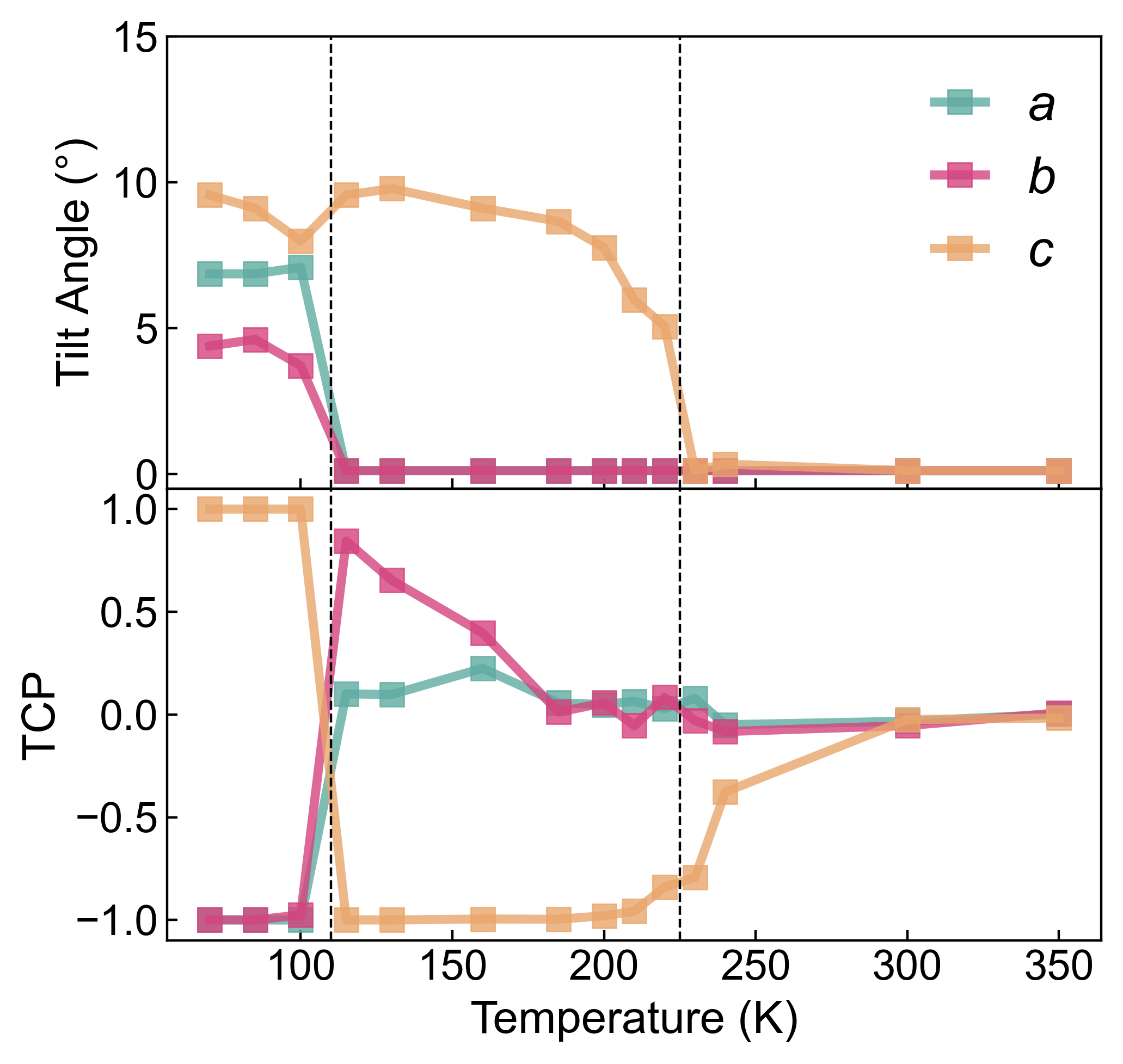}
    \caption{Octahedral tilt angle (upper panel) and tilting correlation polarity (lower panel) of \ce{MAPbBr3} versus temperature. }
    \label{res:tilt_T}
\end{figure}

\hl{The predicted $\alpha$-to-$\beta$ transition temperature (225$\,$K) is well reproduced, and that of $\beta$-to-$\gamma$ transition (110$\,$K) is underestimated by roughly 40$\,$K.} 
This determination of phase transition temperature is achieved by performing multiple constant-temperature NpT MD simulations and finding the temperature range in which the tilting pattern has changed (Fig.~\ref{res:tilt_T}). 
For the octahedral distortion magnitudes, the dynamic distortion (Fig.~\ref{res:gaussians}d) will increase with temperature due to thermal agitation. This is opposite to the distortions of the \textit{time-averaged} crystal structure, which decrease with increasing temperature because of the higher global symmetry of the structure. The distortion magnitudes cannot uniquely distinguish between the crystallographic phases, but the information contained in the octahedral distortion has the potential for featurisation covering multiple perovskite compositions, as different materials may manifest characteristic distortion patterns. 

\begin{figure*}[htb]
    \centering    \includegraphics[width=0.85\textwidth]{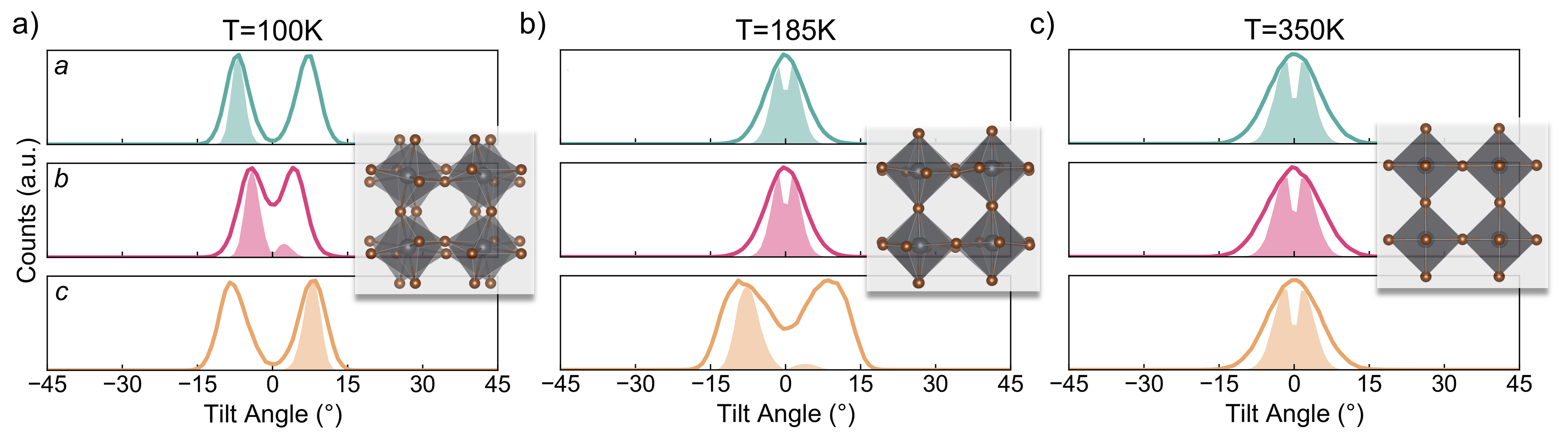}
    \caption{Octahedral tilting in a) orthorhombic phase at 100$\,$K, (b) tetragonal phase at 185$\,$K, (c) cubic phase at 350$\,$K of \ce{MAPbBr3}. Each panel corresponds to one axis. The solid lines denote the dynamic distribution of tilting. The shaded area below the solid lines is the correlation of the tilt angle with the next nearest neighbour along the same direction, which is equivalent to the histogram of $R_{\alpha,\alpha}(k)$. The corresponding global Glazer tilting pattern is $a^-b^-c^+$, $a^0b^0c^-$, and $a^0a^0a^0$, respectively.}
    \label{res:tilt}
\end{figure*}

A core functionality of \textsc{PDynA} is the extraction of octahedral tilting status by isolating the distortion effect. We can directly calculate the octahedral tilt angles and the first nearest neighbour correlation of tilting. 
This analysis can assign the Glazer notation~\cite{glazer_original} of the equilibrated structure, which acts as a straightforward indicator of the crystallographic phase. 
For the orthorhombic phase of \ce{MAPbBr3} shown in Fig.~\ref{res:tilt}a, all three axes exhibit non-zero tilting, and an in-phase tilting pattern is observed from the $c$-axis panel (the shaded peak in the positive side is significantly larger than that of the negative side). The corresponding Glazer notation for the orthorhombic phase is $a^{+}b^{-}c^{-}$ ($P2_{1}/m$), and the $a^{+}$ component refers to the in-phase tilting correlation in the $c$-axis. The remaining $b^{-}c^{-}$ components indicate anti-phase correlation along $a$ and $b$, which have distinct octahedral tilting magnitudes. This is different from the $a^{+}b^{-}b^{-}$ ($Pnma$) symmetry of \ce{MAPbI3}.~\cite{mapi_exp_temp} 
Thus, the tilting correlation polarity (TCP) values of this phase are approximately $[-1,-1,1]$. 

The tetragonal phase shown in Fig.~\ref{res:tilt}b exhibits matches its Glazer notation $a^{0}a^{0}c^{-}$. The octahedral tilting about $a$ and $b$ give single peaks centred at zero degrees, and the $c$-axis has an approximate 12$^{\circ}$ anti-phase tilting with TCP values of $[0,0,-1]$. Lastly, Fig.~\ref{res:tilt}c illustrates the tilting pattern of the equilibrated cubic phase, where all three axes possess zero-centred tilting (with a Glazer notation of $a^{0}a^{0}a^{0}$). Moreover, all zero-centred tilting peaks ($a$ and $b$-axis in the tetragonal phase, and all three axes in the cubic phase) have separated positive and negative correlation functions. This counter-intuitive phenomenon can be explained by the fact that even though the absolute tilting has a statistical mean of zero, the octahedra still instantaneously tilt. Nearest neighbours tilt to approximately the same angle, and this has no preferred correlation polarity (in-phase or anti-phase). Similar to the distortion and lattice parameter quantification, for each temperature, three numbers can define the global tilting pattern by performing Gaussian fitting on the absolute tilting angle distribution, and another three for the computed TCP values.

The role of temperature is shown in Fig.~\ref{res:tilt_T}, with changes in the tilt angles (upper panel) and the corresponding TCP values (lower panel). All three principal axes of the cubic phase have zero-centred tilt angles, one of which becomes finite in the tetragonal phase (taking values from 5$^{\circ}$ to 10$^{\circ}$ according to the temperature), and the orthorhombic phase exhibits three distinguished non-zero tilt angles. It is also found that both phase transition processes of \ce{MAPbBr3} are first-order transitions. The TCP values, however, can further reveal the spatial relations behind the octahedral tilt angles. 

For the higher temperature regime of each phase, TCP values are consistent with the formal Glazer notation. As phase transition temperatures approached, peculiar phenomena were found. At 240$\,$K, even though the tilt angles remain in the cubic form, the TCP value of one of the axes is driven away from zero, similar to that of the tetragonal phase. This means that the first nearest neighbours in the $c$ direction tend to have anti-phase correlation rather than in-phase correlation (the Glazer plot for this temperature is shown in Fig.~S1a). 
Likewise, at 130$\,$K close to the $\beta-\gamma$ transition, not only the tetragonal $c$-axis but also one of the other two zero-tilting axes exhibit uneven correlation. 
In this case, the $b$-axis possesses an in-phase correlation with a zero-centred tilting (shown in Fig.~S1b). Both phenomena imply that as the cubic and tetragonal phases are cooled towards the phase transition, this dynamic tilting effect will occur and lower the effective crystal symmetry.

\subsubsection{Molecular orientation: distribution, spatial correlation and dynamics}
\begin{figure*}[htb!]
    \centering   \includegraphics[width=0.7\textwidth]{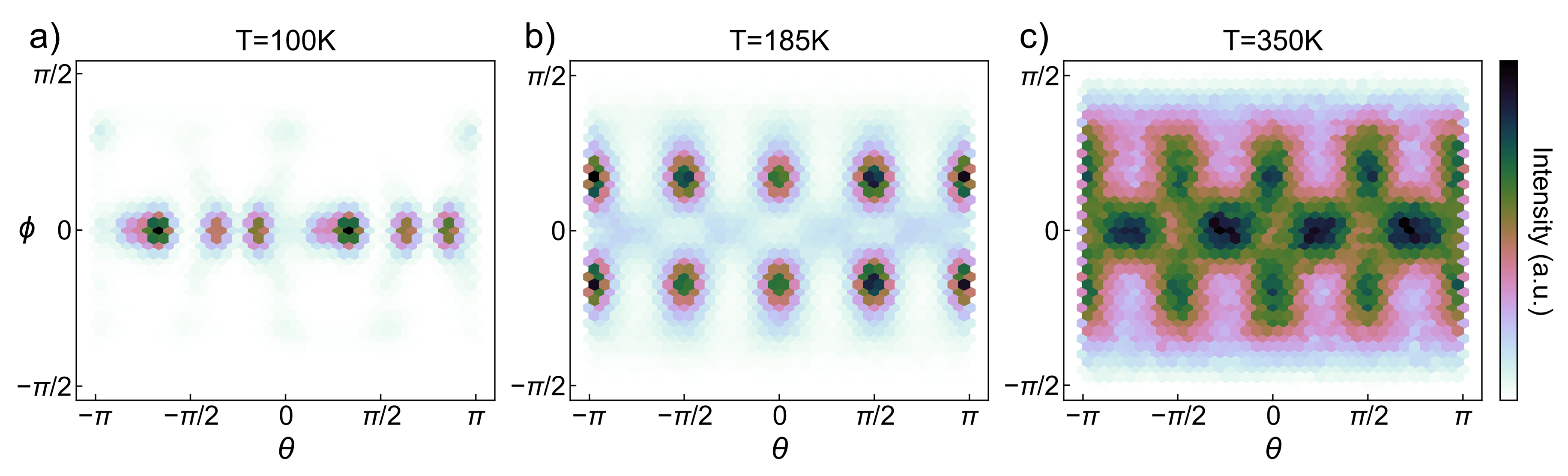}
    \caption{Molecular orientation (MO) distribution in spherical coordinates of (a) orthorhombic phase at 100$\,$K, (b) tetragonal phase at 185$\,$K, (c) cubic phase at 350$\,$K of \ce{MAPbBr3}. The orientations of the A-site molecule are projected onto the horizontal (azimuthal angle $\theta$) and vertical (polar angle $\phi$) axes. }
    \label{res:MO}
\end{figure*}

In contrast to the strongly interconnected octahedral network, the molecular components have more orientational freedom. Each crystallographic phase has a characteristic preferred molecular orientation (MO), as illustrated in Fig.~\ref{res:MO}. 
MA ions in the orthorhombic phase (Fig.~\ref{res:MO}a) are parallel to the $ab$-plane and are confined in several local minima. 
A different pattern is found in the tetragonal phase (Fig.~\ref{res:MO}b) where the molecules adopt eight clear preferred orientations that are symmetrical about the $ab$-plane, each forming an angle of approximately 31.4$^{\circ}$ with the $ab$-plane. The MO of the cubic phase (Fig.~\ref{res:MO}c) has a much broader distribution. The MO is almost evenly spread in all directions except for those towards the nearest Pb atoms. 
The spherical coordinate mapping underestimates the population near the poles, especially for the cubic phase. The normalised 3D visualisation is shown in Fig. S2.  

The molecular alignment is also affected by the adjacent molecules as quantified by the MO spatial correlation functions. 
Here we will focus on the first and second nearest neighbours to reconcile with the repeating pattern of octahedra. 
In the orthorhombic phase the first nearest neighbour correlation function in the $a$ direction $C_{a}^{(1)}$ (the shaded curve in the $a$ panel of Fig.~S4a) has two main populations centred at \textit{ca.} 0.8 and -1. This means that the angle between two adjacent molecules in the $a$ direction is likely to be either 37$^{\circ}$ or 180$^{\circ}$. Interestingly, $C_{c}^{(1)}$ (the shaded curve in the $c$ panel of Fig.~S4a) has a mean value close to -1 and that of $C_{c}^{(2)}$ (the solid line) is 1, implying that the molecules are alternatively aligned in the $c$ direction. The analysis is similar for the tetragonal and cubic phases. As the temperature is increased, the molecules become less correlated in space. Moreover, these correlation functions directly reflect the symmetry of the structure. In the orthorhombic phase, the molecules are correlated differently in each direction, but the symmetry relations of $a=b$ and $a=b=c$ are obeyed for the tetragonal and cubic phases.

These correlation functions contain higher-dimensional information that cannot be utilised as scalar descriptors. Instead, we introduce two order parameters, namely the alignment factor (AF) and contrast factor (CF), listed in Table~\ref{corr}. 
AF measures, for the adjacent neighbours in each direction, the difference between amounts of aligned and anti-aligned molecules. For example, AFs of the orthorhombic phase all take different values, but for the tetragonal phase in Fig.~S4b only AF$_{c}$ is distinct. 
However, this metric does not distinguish between the tetragonal and cubic phases. Instead, the CF compares the first and second nearest neighbour MO correlation functions. From Fig.~S4, CF should have the lowest value (least similar) in the orthorhombic phase and the highest value (almost identical) in the cubic phase.

\begin{table}[ht]
\centering
\caption{Characteristic molecular orientation (MO) order parameters for each crystallographic phase of \ce{MAPbBr3} 
including the alignment factor (AF) and contrast factor (CF). Values for a 1D rod of aligned and anti-aligned molecules are shown for comparison. 
}
\begin{tabular}[t]{c@{\hskip 0.3in}c@{\hskip 0.2in}c@{\hskip 0.2in}c@{\hskip 0.2in}c}
Phases & \multicolumn{4}{c}{MO order parameters} \\
\hline
 & $AF_{a}$ & $AF_{b}$ & $AF_{c}$ & $CF$ \\
$[\uparrow \uparrow \uparrow \uparrow \cdots]$ & 1.0 & & & 1.0 \\
$[\uparrow \downarrow \uparrow \downarrow \cdots]$ & -1.0 & & & -1.0 \\
$\alpha$, 350K & -0.05& -0.05& -0.05& 0.9 \\
$\beta$, 185K & 0.05& 0.05& -0.25& 0 \\
$\gamma$, 100K & -0.40& 0.75& -1& -0.75 \\
\hline
\end{tabular}
\label{corr}
\end{table}%

\begin{figure}[hbt]
    \centering
    \includegraphics[width=0.7\linewidth]{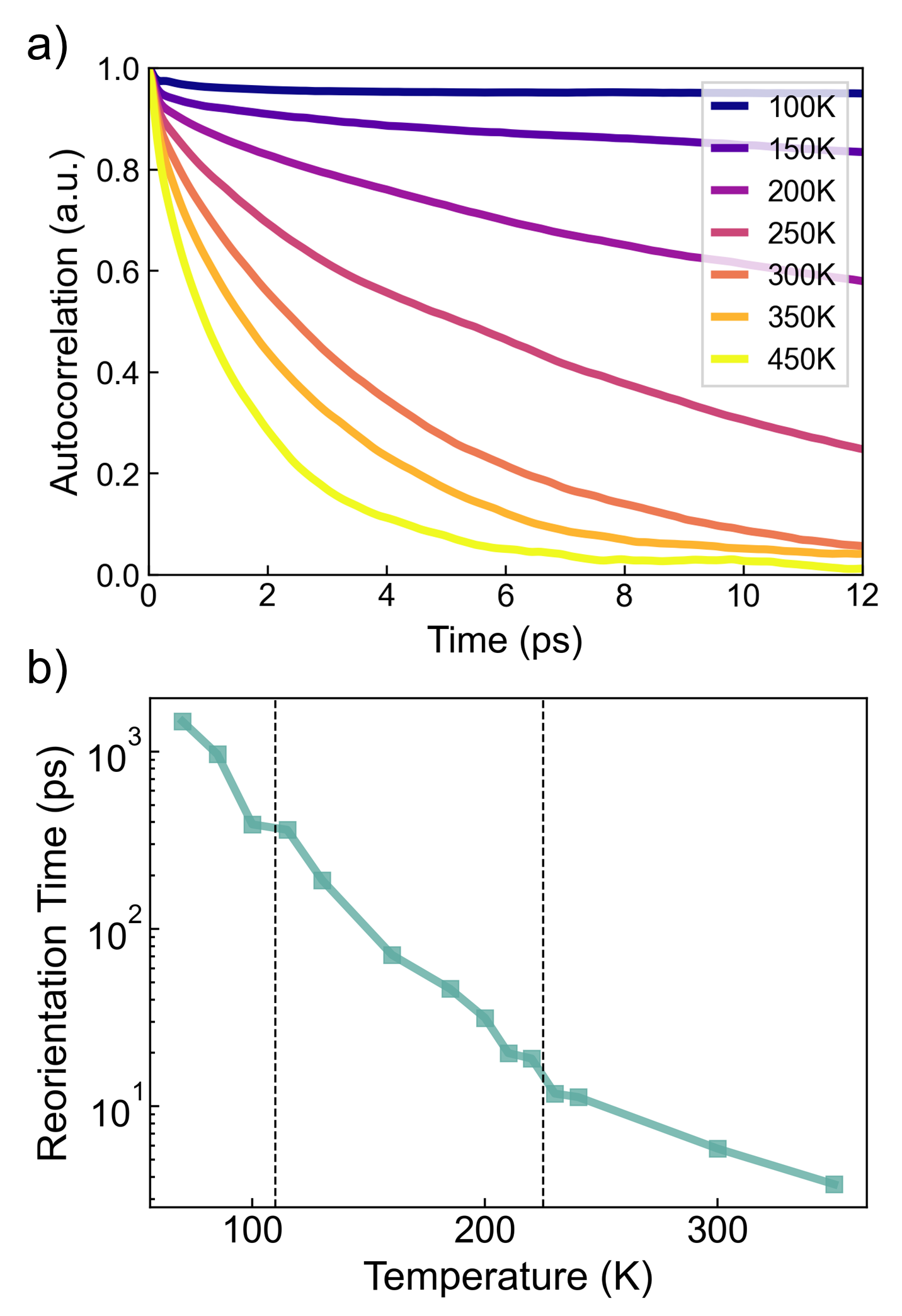}
    \caption{(a) Molecular autocorrelation functions versus time of \ce{MAPbBr3} for multiple temperatures. (b) Molecular reorientation time constants of \ce{MAPbBr3} versus temperature. }
    \label{res:autocorr}
\end{figure}

The A-site molecules are correlated with themselves in time as described by the autocorrelation function, Fig.~\ref{res:autocorr}a. 
At lower temperatures, the molecule tends to freeze in one orientation so that the autocorrelation function stays close to 1. Higher temperatures allow larger amplitude vibrations and hops between minima. 
The decay of the autocorrelation function is fitted with the exponential function with a characteristic decay time constant (see Methods). 
The time constants surge rapidly when the temperature is reduced, and are relatively independent of the crystallographic phases. Here we obtained molecular reorientation dynamics that are approximately an order of magnitude faster than calculations made by Jinnouchi \textit{et al.}~\cite{vaspmlff_original}. 
We attribute this to their methodology that artificially increases the H mass, which hinders the molecular rotational dynamics. The calculated room temperature reorientation time of the MA molecule is approximately 7$\,$ps, in good agreement with other calculations of 4--15 ps\cite{mapbbr_sim2016even} and experimental measurements of 3--14$\,$ps~\cite{mapbx_exp2017sarma}. 

\subsubsection{Transient Properties}
\begin{figure}[htb]
    \centering
    \includegraphics[width=0.8\columnwidth]{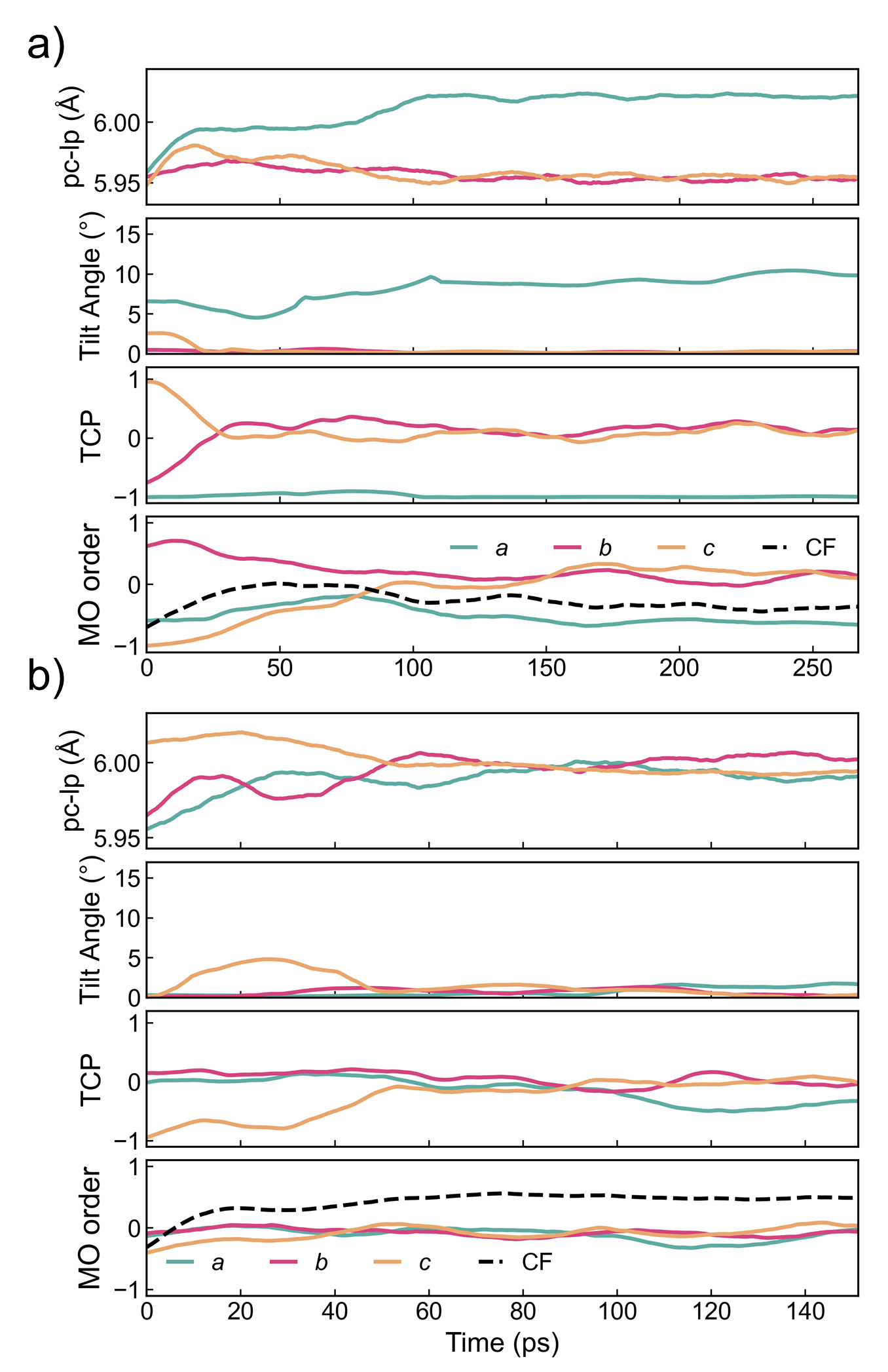}
    \caption{Transient structural equilibration of simulated local lattice parameters (pc-lp), octahedral tilting, tilting correlation polarity (TCP) and molecular orientation (MO) order descriptors (three alignment factors and a contrast factor). (a) Orthorhombic phase equilibrated at 80$\,$K and run at 170$\,$K, (b) tetragonal phase equilibrated at 150$\,$K and run at 240$\,$K of \ce{MAPbBr3}.}
    \label{res:transient}
\end{figure}

Apart from the equilibrium properties calculated from each temperature, one would also be interested in quantifying the non-equilibrium dynamics of transitions between perovskite phases. 
This can be achieved by introducing a temperature gradient or by changing the temperature of an equilibrated phase. 
We utilised the latter approach due to the slow A-site rotational dynamics. 
A structure was first equilibrated with constant-temperature NpT MD and then the atomic positions and velocities of the last frame in this calculation are used as the initial configuration of a second NpT MD, where the temperature is set to another temperature above the phase transition. 
In both orthorhombic--tetragonal and tetragonal--cubic phase transitions shown in Fig.~\ref{res:transient}, a continuous evolution of all three properties can be observed, where the corresponding properties change progressively. At 170$\,$K, the TCP values, or equivalently the pattern of coordinated tilting mode of octahedra will first converge to the tetragonal form within 50$\,$ ps, Fig.~\ref{res:transient}a. Then the octahedral tilt angles and local lattice parameters (\textit{i.e.} relative spacing between octahedra) will converge within 100$\,$ ps. This can be rationalised by the fact that tilting directly dictates the spacing between octahedra. Lastly, the molecules adapt their tetragonal form after 150$\,$ ps of heating. Likewise for the tetragonal--cubic transition at 240$\,$K (Fig.~\ref{res:transient}b), all four properties transformed into their cubic form in 40 to 60$\,$ ps. Less information in this process can be extracted due to a more rapid transition, but we do observe that the reorientation of the organic A-site molecules is driven by octahedral tilting. We note that the reverse process of these two transitions is not accessible, as directly cooling the structures to a lower temperature will not trigger phase transition (the required time span for this process is longer than the allowed simulation time). This is because the kinetics of atoms at a lower temperature is slower, and the A-site molecules possess a freezing effect that impede phase transition upon cooling.~\cite{vaspmlff_perov}

\begin{figure*}[!htb]
    \centering
    \includegraphics[width=0.74\textwidth]{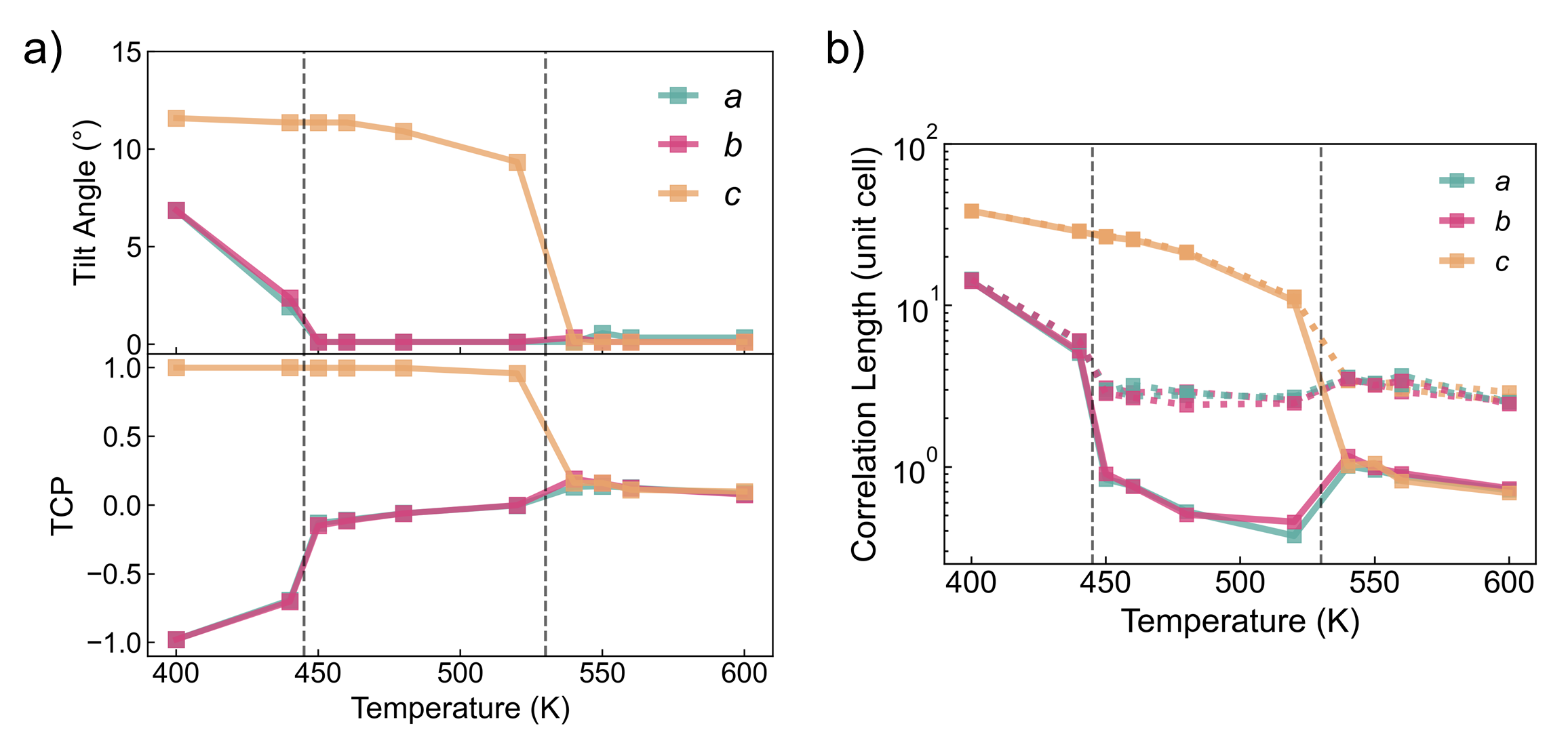}
    \caption{Properties of \ce{CsPbI3} versus temperature, including (a) octahedral tilt angles and TCP values, (b) correlation length of tilting in space, where the solid lines are diagonal terms of the spatial correlation tensor ($R_{\alpha,\beta}(k)$ with $\alpha=\beta$) and the dashed lines are the off-diagonal terms ($R_{\alpha,\beta}(k)$ with $\alpha \neq \beta$). }
    \label{res:cs1}
\end{figure*}

\subsection{Analysis of Large Scale \ce{CsPbI3} Trajectories}

To test the capability of \textsc{PDynA}, we analysed MD trajectories of \ce{CsPbI3} structures with a supercell size of $24\times24\times24$ that contains 13,824 octahedra. Since the A-site of this material is inorganic, we focus on the behaviour of the octahedral network.

The crystallographic phases found, from left (400 K) to right (600 K), are orthorhombic ($a^{+}b^{-}b^{-}$), tetragonal ($a^{0}a^{0}c^{+}$), and cubic ($a^{0}a^{0}a^{0}$), Fig.~\ref{res:cs1}a.
The corresponding titling distributions are shown in Figure S5.
Within the cubic phase, as the temperature is lowered towards the phase transition, all three TCP values increase slightly, implying that all axes exhibit subtle in-phase correlation with their neighbours. This differs from the tilting correlation behaviour of \ce{MAPbBr3}, where only one axis possesses an unequal TCP value. 
Local tilt domains favour in-phase correlation along all three axes of \ce{CsPbI3}. 
The size of these domains can also be computed through the correlation length $\xi$ of tilting in space, Fig.~\ref{res:cs1}b. 

For the cubic phase, the three diagonal and six off-diagonal contributions are isotropic with respect to the principal axes. This leaves two distinct values in the correlation tensor. 
The off-diagonal term $\xi_{\alpha,\beta}$ is approximately 3 unit cells long, so octahedral tilting in one direction will correlate up to its third nearest neighbour in the other two directions. 
In contrast, the diagonal term $\xi_{\alpha,\alpha}$ is mostly under 1 unit cell length, indicating the absence of correlated titling even for the first nearest neighbour. 
Combining these two observations, the local tilt domain in the cubic phase has a shape analogous to a plate, which is roughly 1 unit cell thick and has a radius of 3 unit cells long. 
This analysis is consistent with the plate-like domains suggested from neutron scattering of \ce{CsPbBr3} by Lanigan-Atkins et al\cite{overdamped2021delaire}, which have also been found in
\ce{CH3NH3PbI3}\cite{weadock2023nature,tilt2023erhart}.

\subsubsection{3D Property Maps}

\begin{figure*}[hbt]
    \centering
    \includegraphics[width=0.8\textwidth]{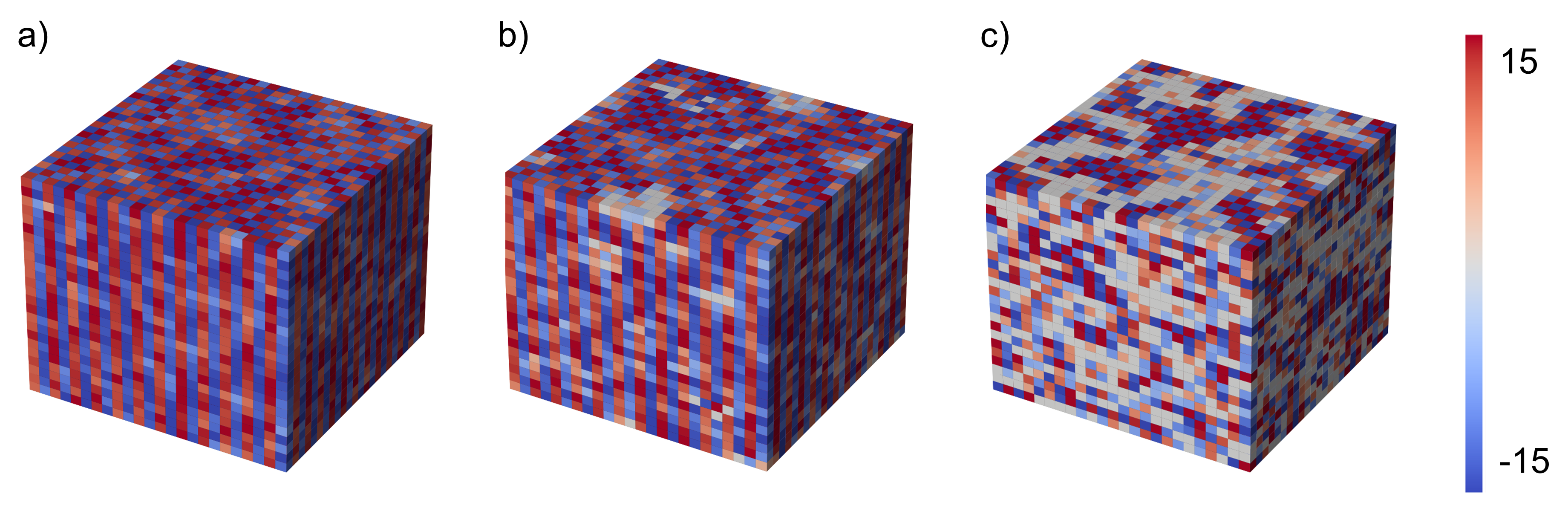}
    \caption{Representation of the $c$ tilt angle of one frame in a dynamic trajectory. Each voxel corresponds to an octahedron in the structure of \ce{CsPbI3}. The $c$ axis is pointing up and the colour bar maps between the most positive and negative tilt angles. (a) Orthorhombic phase at 400$\,$K, (b) tetragonal phase at 480$\,$K, (c) cubic phase at 600$\,$K, where on the top (001) plane, the tilt angles of neighbouring octahedra in a local domain form a chequerboard (+/-) pattern. On the side planes, these domains are sliced to reveal their disc-like nature.}
    \label{sm:cs3D}
\end{figure*}

The spatial arrangement of structural features can also be plotted and analysed directly. For this we employ the voxel module in \textsc{matplotlib}.

The instantaneous octahedral tilting angles are visualised for a structural snapshot of the three \ce{CsPbI3} phases in Fig.~\ref{sm:cs3D}.
The locked titling of the orthorhombic phase is clearly observed in this single snapshot, consistent with the global $a^{-}b^{-}\mathbf{c^{+}}$ pattern.
The tetragonal phase shows similarly locked $c$ tilting consistent with its $a^{0}a^{0}\mathbf{c^{+}}$ pattern; however, small regions of zero tilting are found.
In contrast, high disorder can be seen in the cubic phase (average $a^{0}a^{0}\mathbf{a^{0}}$ pattern) with the formation of local plate-like tilting domains. 
These plots illustrate octahedral tilts along the $c$ axis only and correlated tilting may occur simultaneously along the other axes.

We note that the correlation length metric $\xi$ (Fig. \ref{res:cs1}b) does not take into account in-phase and anti-phase correlation effects; both modes occur with similar probability. In the tetragonal phase, a strong correlation of the $c$ tilt is found in all three directions, while the $a$ and $b$ tilts possess weaker spatial correlations as they are in fact pure $a^{0}$ axes. Lastly, the orthorhombic phase has strongly correlated tilting in all directions, consistent with the condensation of the corresponding soft phonon modes.

\section{Conclusion}
We have defined and implemented a set of \hl{compact} structural descriptors for analysing the average and local structure of perovskite crystals.
The examples of \ce{CH3NH3PbBr3} and \ce{CsPbI3} were chosen with large MD trajectories produced using machine learning force fields that were trained from \textit{ab initio} data. 
Importantly, the global features of each perovskite polymorph were reproduced including structural anisotropy and octahedral tilting patterns. 
We gained further insights into the local correlations within the inorganic and molecular sublattices, including transient domain formation. 
The descriptors and systematic analysis framework are general and can be transferred to more compositionally and structurally complex perovskites.

~

\section{Supporting Information} 


A set of additional analysis plots. Figure S1: illustration of local lattice parameters. Figure S2: octahedral tilting of \ce{MAPbBr3} close to phase transition temperatures. Figure S3: three-dimensional visualisation of MO of \ce{MAPbBr3}. Figure S4: fitting of MO autocorrelation function. Figure S5: MO correlation function of \ce{MAPbBr3}. Figure S6: octahedral tilting of \ce{CsPbI3}.

~

\section{Code Availability}
The \textsc{PDynA} package developed in this work is open-source and available online at \url{https://github.com/WMD-group/PDynA} (DOI: 10.5281/zenodo.7948045)

\section{Acknowledgements}

We thank A. M. Ganose, S. R. Kavanagh, J. M. Frost, K. Tolborg, M. Dubajic, A. Iqbal, K. Morita, L. Zeng and X. Fan for helpful discussions. J. K. acknowledges support from the Swedish Research Council (VR) program 2021-00486. Via our membership of the UK's HEC Materials Chemistry Consortium, which is funded by EPSRC (EP/X035859/1), this work used the ARCHER2 UK National Supercomputing Service (http://www.archer2.ac.uk).

\bibliographystyle{apsrev4-1}
\bibliography{REF.bib}

\end{document}


\title{Supporting Information for ``Structural Dynamics Descriptors for Metal Halide Perovskites"}

\author{Xia Liang}
\affiliation{Department of Materials, Imperial College London, South Kensington Campus, London SW7 2AZ, UK\\}

\author{Johan Klarbring}
\affiliation{Department of Materials, Imperial College London, South Kensington Campus, London SW7 2AZ, UK\\}
\affiliation{Department of Physics, Chemistry and Biology (IFM), Link\"{o}ping University, SE-581 83, Link\"{o}ping, Sweden}

\author{William Baldwin}
\affiliation{Department of Engineering, University of Cambridge, Cambridge CB2 1PZ, UK}

\author{Zhenzhu Li}
\affiliation{Department of Materials, Imperial College London, South Kensington Campus, London SW7 2AZ, UK\\}

\author{G\'abor Cs\'anyi}
\affiliation{Department of Engineering, University of Cambridge, Cambridge CB2 1PZ, UK}

\author{Aron Walsh}
\email{a.walsh@imperial.ac.uk}
\affiliation{Department of Materials, Imperial College London, South Kensington Campus, London SW7 2AZ, UK\\}
\affiliation{Department of Physics, Ewha Womans University, Seoul 03760, Korea}

\maketitle

\beginsupplement

\begin{figure}[h]
    \centering
    \includegraphics[width=0.7\linewidth]{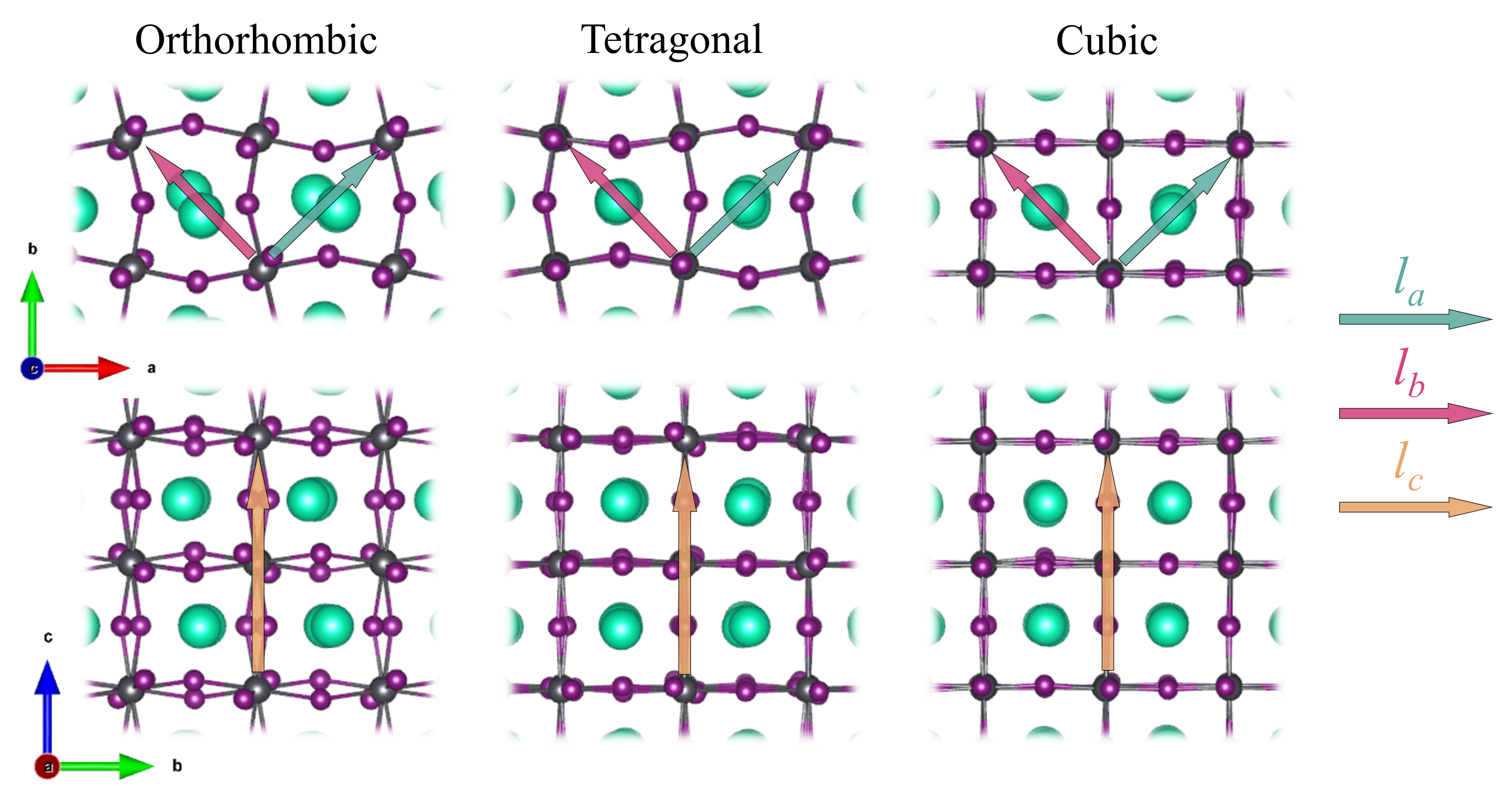}
\caption{
Illustration of the local lattice spacing in three perovskite phases (orthorhombic, tetragonal, and cubic) defined with respect to the relative positions of the B-site cations.}
    \label{pseudocubic}
\end{figure}

\begin{figure}[htb]
    \centering
    \includegraphics[width=0.7\linewidth]{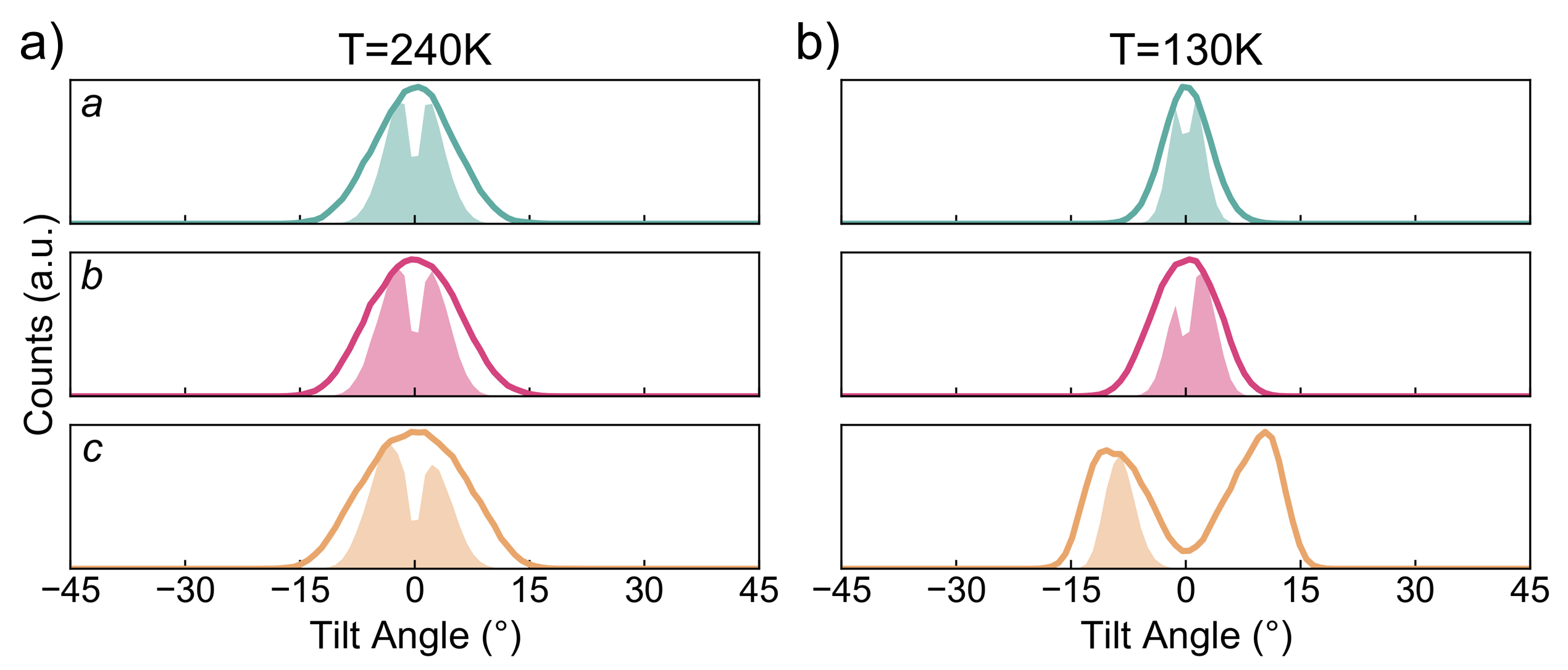}
    \caption{Octahedral tilting at (a) 240$\,$K and (b) 130$\,$K in \ce{MAPbBr3}.}
    \label{sm:240K_130K}
\end{figure}

\begin{figure*}[htb]
    \centering
    \includegraphics[width=0.7\textwidth]{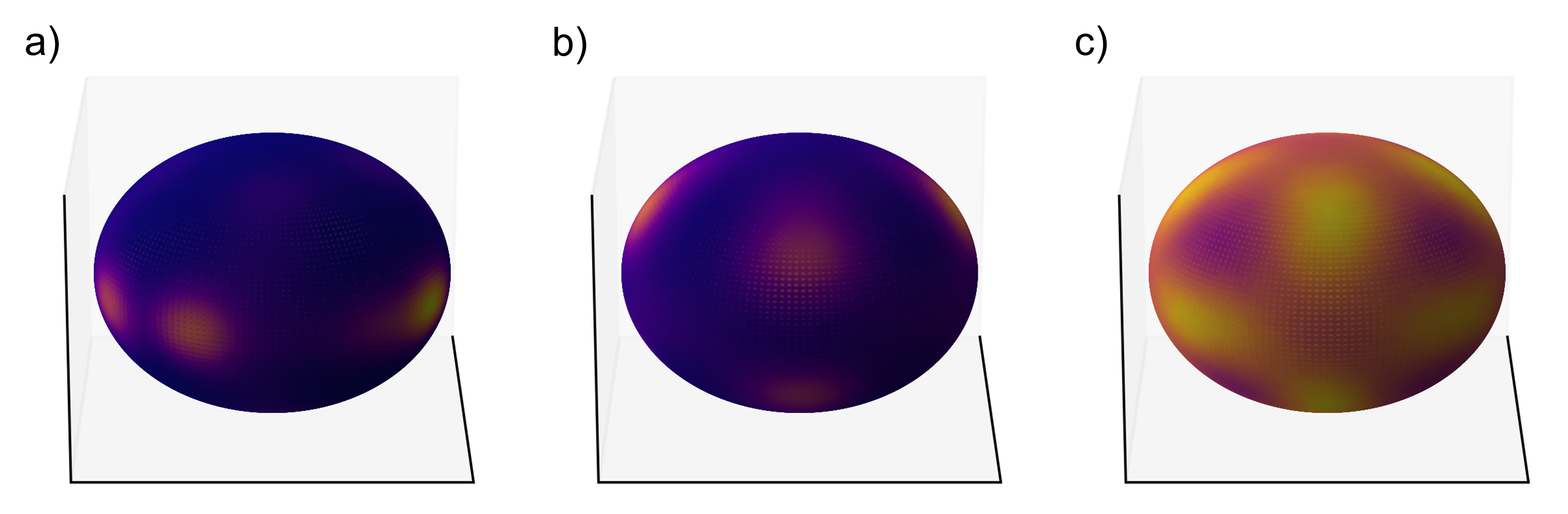}
    \caption{Molecular orientation distribution visualised in three-dimensional space, showing the same quantity as in Fig. 6 of the main text. (a) Orthorhombic phase at 100$\,$K, (b) tetragonal phase at 185$\,$K, (c) cubic phase at 350$\,$K of \ce{MAPbBr3}. }
    \label{sm:MO3D}
\end{figure*}

\begin{figure}[htb]
    \centering
    \includegraphics[width=0.4\linewidth]{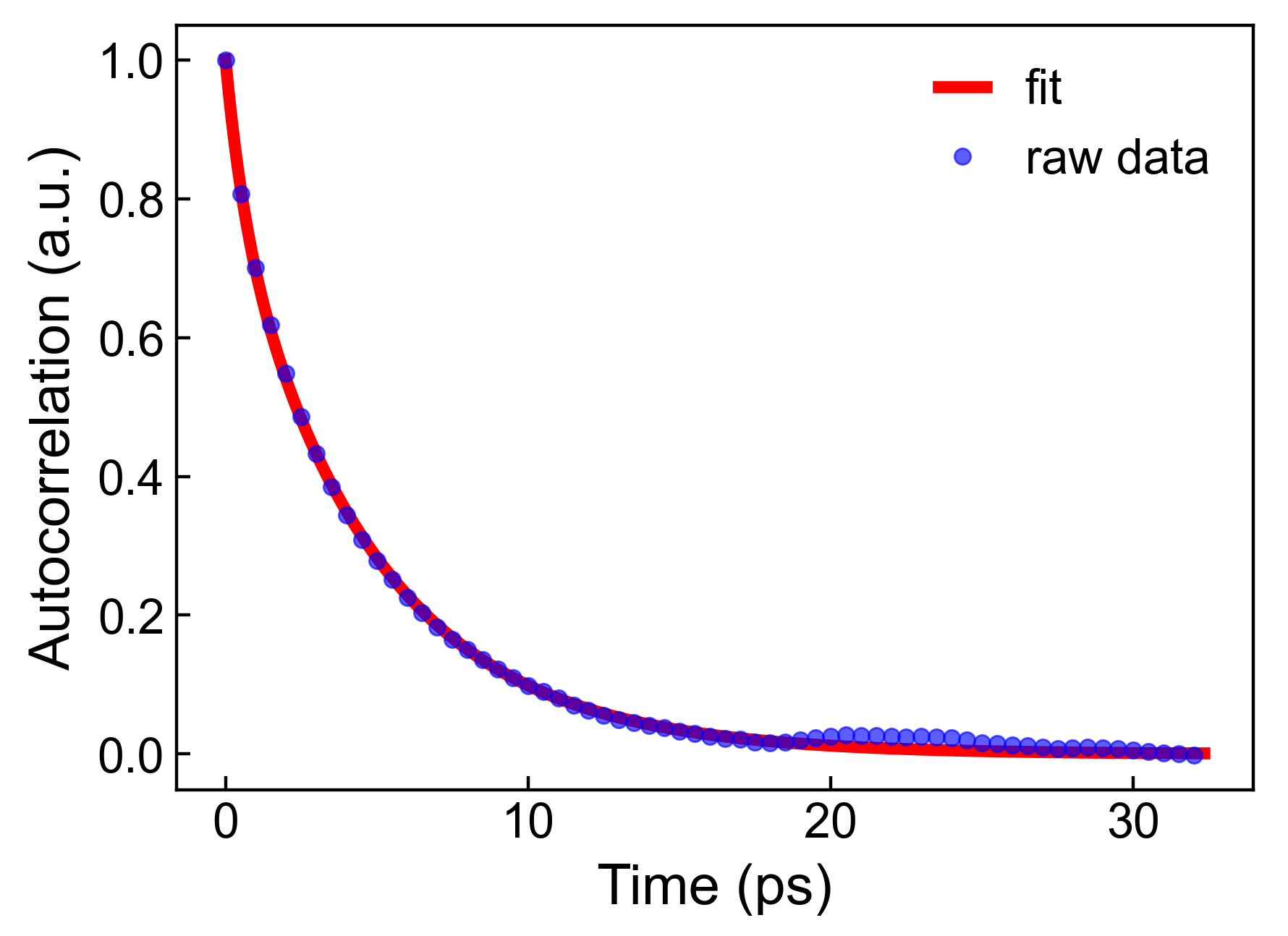}
    \caption{Illustration of the fitting of molecular orientation autocorrelation function $A_{MO}(t)$ for \ce{CH3NH3}. }
    \label{sm:autocorr}
\end{figure}

\begin{figure*}[htb]
    \centering
    \includegraphics[width=0.76\textwidth]{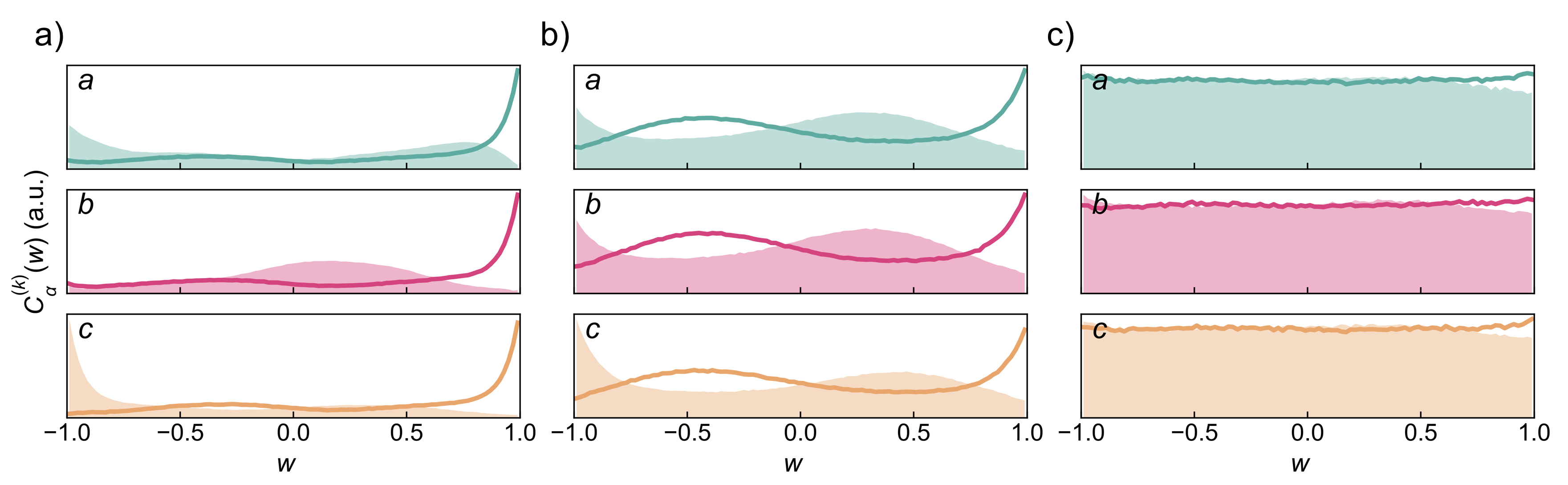}
    \caption{Molecular orientation correlation function of \ce{MAPbBr3} with (a) orthorhombic phase at 100$\,$K, (b) tetragonal phase at 185$\,$K, (c) cubic phase at 350$\,$K. The shaded population in each panel refers to the first nearest neighbour correlation function $C_{\alpha}^{(1)}$ and the solid line is the second nearest neighbour counterpart $C_{\alpha}^{(2)}$. }
    \label{res:MOcorr}
\end{figure*}

\begin{figure*}[htb]
    \centering
    \includegraphics[width=0.84\textwidth]{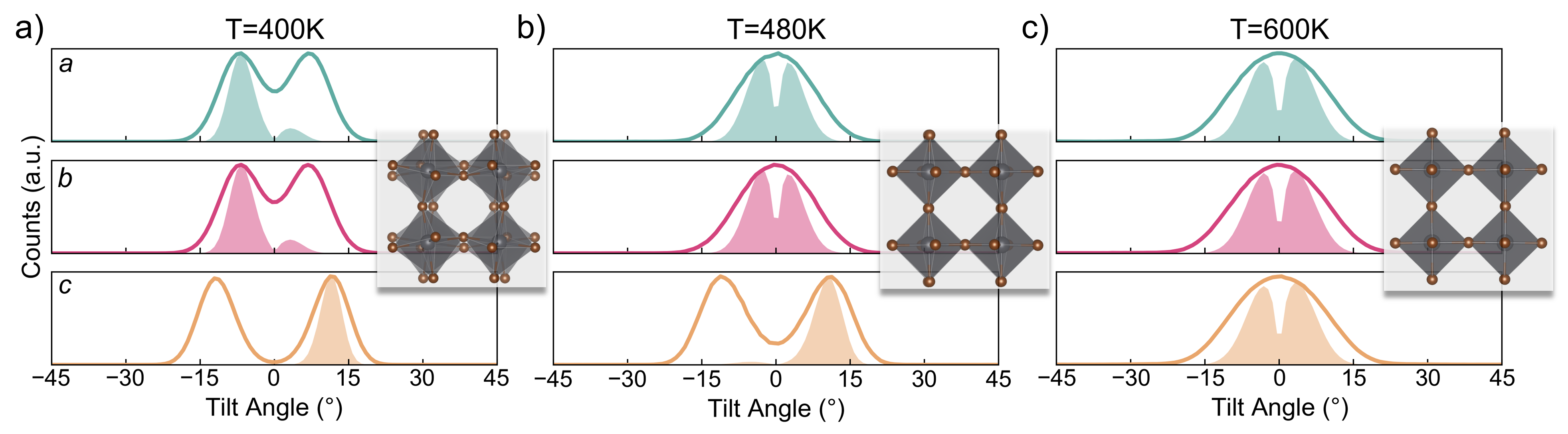}
    \caption{Octahedral tilting in (a) orthorhombic phase at 400$\,$K, (b) tetragonal phase at 480$\,$K, (c) cubic phase at 600$\,$K of \ce{CsPbI3}.}
    \label{sm:cs}
\end{figure*}

\begin{figure*}[htb]
    \centering
    \includegraphics[width=0.54\textwidth]{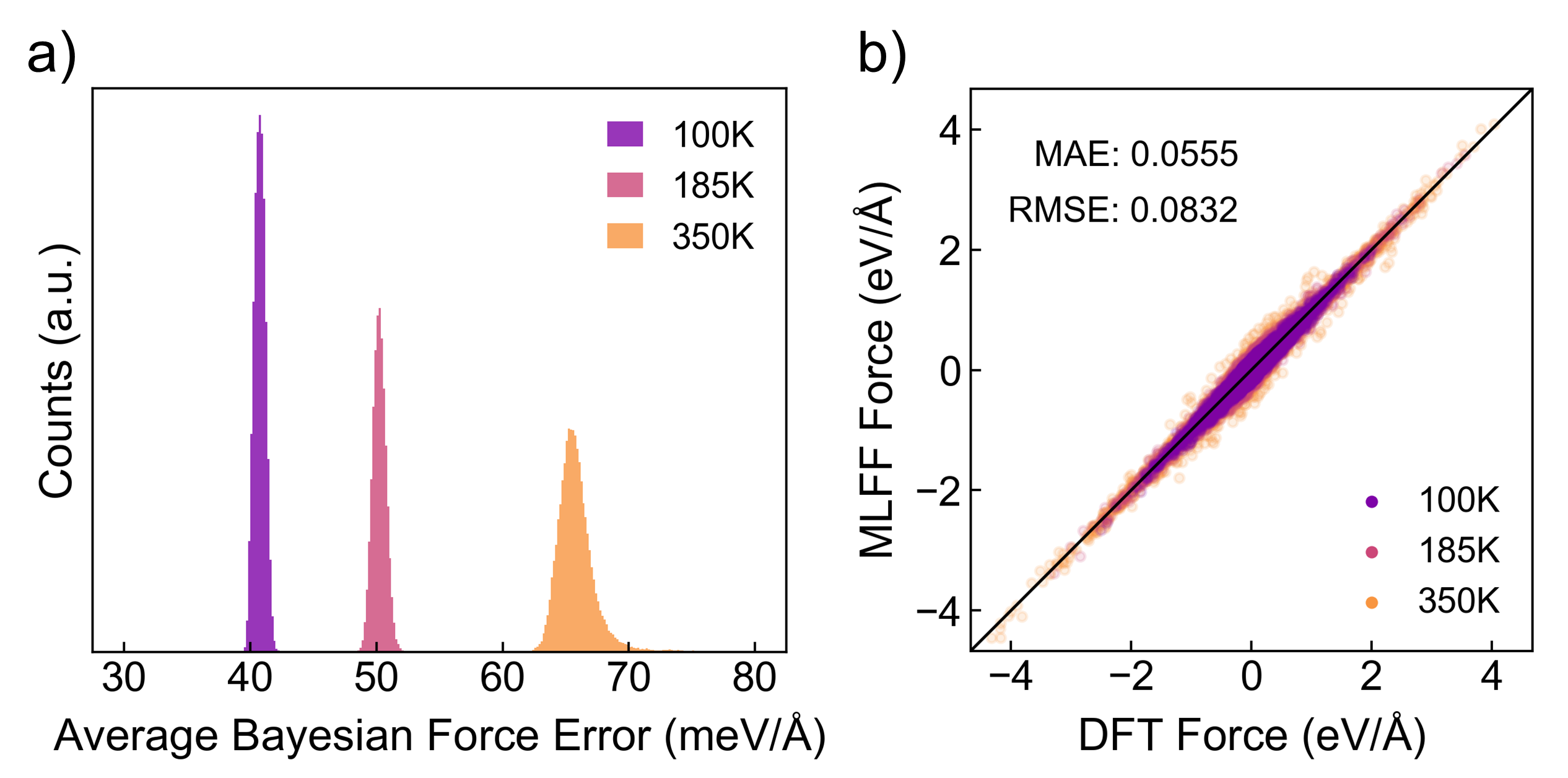}
    \caption{(a) MLFF Bayesian error estimation of force per atom, and (b) benchmark between DFT forces and MLFF forces on each atom of \ce{MAPbBr3} at 100$\,$K, 185$\,$K and 350$\,$K. Three MLFF MD calculations are performed on a $4\times4\times4$ supercell at the above temperatures. Five frames are randomly selected from each trajectory. The DFT forces are calculated on these frames with the same settings as the MLFF training step, which are compared to the corresponding MLFF forces. }
    \label{sm:error}
\end{figure*}